\documentclass{desyproc}
\begin{document}
\title{Search for Hidden photons with Sumico}

\author{{\slshape Yoshizumi Inoue$^1$,
    Tetsuya~Mizumoto$^2$, Ryosuke~Ohta$^1$,
    Tomoki~Horie$^1$, Jun'ya~Suzuki$^1$,\\ Makoto~Minowa$^1$}
  (the Sumico collaboration)\\[1ex]
  $^1$The University of Tokyo, Japan\\
  $^2$Kyoto University, Japan}

\contribID{inoue\_yoshizumi}

\desyproc{DESY-PROC-2013-04}
\acronym{Patras 2013} 
\doi  

\maketitle

\begin{abstract}
We searched for solar hidden photons in the visible photon energy range
using a hidden photon detector add-on attached to Sumico.
It consists of a parabolic mirror of $\phi0.5\rm\,m$ and $f=1\rm\,m$
installed in a vacuum chamber,
and a low noise photomultiplier tube at the focal point.
No evidence for the existence of hidden photons was found
in the latest measurement
giving a new limit on the photon-hidden photon mixing parameter
in the hidden photon mass range 0.001--1\,eV.
\end{abstract}

\section{Introduction}

The sun can be a powerful source of weakly interacting light particles,
such as the axions~\cite{PQ1,PQ2}.
The Tokyo axion helioscope, aka. Sumico~\cite{Sumico,Sumico1,Sumico2,Sumico3},
was built aiming at the direct detection of the solar axions
in the mass range up to a few eV.
It is equipped with a dedicated cryogen-free superconducting magnet
which can produce a transverse magnetic field of 4\,T over 2.3\,m,
a container to hold cold $^4$He gas,
a PIN-photodiode-array X-ray detector,
and a telescope mount mechanism to track the sun.
The transverse magnetic field is essential in the axion helioscope,
where the solar axions oscillate into X-ray photons.
In the past measurements, the axion mass ranges 0--0.27\,eV and 0.84--1\,eV
have been scanned.
We were striving to push up the sensitive mass range to higher masses.
Unfortunately, however, axion search activity is currently suspended
due to a trouble in the cryogenic system
which is preventing the magnet from exciting.
Meanwhile, we searched for solar hidden photons using Sumico.

The hidden photon is the gauge boson of a hypothetical hidden local U(1)
symmetry.
Many extensions of the standard model,
in particular those based on string theory,
predict such symmetries~\cite{Ringwald}.
The hidden photon can couple to the ordinary photons
via a so-called \emph{kinetic mixing},
and it can be massive
as described by the following Lagrangian~\cite{Okun1,Holdom,Foot},
\[
  \mathcal{L} = 
  -{1\over4}F_{\mu\nu}F^{\mu\nu}-{1\over4}B_{\mu\nu}B^{\mu\nu}
  -{\chi\over2}F_{\mu\nu}B^{\mu\nu}+{m_{\gamma'}\over2}B_\mu B^\mu,
\]
where $F_{\mu\nu}$ and $B_{\mu\nu}$ represent
the ordinary- and hidden- photon field strengths, respectively,
$\chi$ is the kinetic mixing parameter,
and $m_{\gamma'}$ is the hidden photon mass.
When the hidden photon has a small mass,
it leads to photon--hidden-photon vacuum oscillations.
In vacuum, hidden photon to photon transition probability 
for photons of energy $\omega$ after traveling $\ell$ is given by:
\[
  P_{\gamma'\to\gamma}=4\chi^2\sin^2\left({m_{\gamma'}^2\ell\over4\omega}\right)
\]
assuming $m_{\gamma'}\ll\omega$.
Since the refractive index for visible light in the matter is
normally larger than 1,
the matter in the conversion path always affects the transition probability
negatively.

The emission of hidden photons from the sun
was discussed by J.~Redondo~\cite{Redondo1}.
The transverse hidden photon flux at the earth is given by:
\[
  {{\rm d}\Phi_{\rm T}\over{\rm d}\omega}=
  \int_0^{R_\odot}\!\!{r^2dr\over(1\rm\,AU)^2}\,
  {\omega^2\over\pi^2}\,
  {\Gamma\over e^{\omega/T}-1}\,
  {\chi^2m_{\gamma'}^4\over
  (m_\gamma^2-m_{\gamma'}^2)^2+(\omega\Gamma)^2},
\]
where
$R_\odot$ is the solar radius,
$T$ is the temperature,
$\Gamma$ is the damping rate of photons,
and $m_\gamma$ is the effective photon mass in plasma.
For $m_{\gamma'}\ll1\rm\,eV$,
one can use the following conservative estimate 
for the bulk component of the hidden photon flux~\cite{Redondo1}:
\[
  {{\rm d}\Phi_{\rm T}\over{\rm d}\omega}\gtrsim
  \chi^2\,\left({m_{\gamma'}\over\mathrm{eV}}\right)^4\,10^{32}
  {1\over\rm eV\,cm^2\,s}\qquad\hbox{for}\quad\omega=\hbox{1--5\,eV}.
\]
Recently, a refined estimation
including the contribution from a thin resonant region below the photosphere
is given~\cite{Redondo2,Redondo3}
for four typical cases $m_{\gamma'}=0, 0.01, 0.1$ and $1\rm\,eV$. 
They claim that the resonant production dominates over the emission
from the rest of the sun.

In more recent studies~\cite{An1,Redondo4},
the estimated emission rate of longitudinal hidden photons was revised
and it turned out that the resonant longitudinal-mode production is dominating
the total solar hidden photon emission in the low mass case
$m_{\gamma'}<1\rm\,eV$.
This result has nothing to do with the analysis of our experiment
since longitudinal hidden photons do not oscillate into photons in vacuum.
However, it has a strong impact on the hidden photon searches
by providing stringent constraints on the hidden photon parameters
along with their another work~\cite{An2} in which
they re-analyzed the published data of the XENON10 dark matter experiment.

Various constraints on the hidden photon parameters are summarized in
Ref.~\cite{JaeckelRingwald}.

In this paper, we report on
a direct experimental search for the flux of solar hidden photons.

\section{Experimental apparatus}

\begin{wrapfigure}[13]{r}{0.5\textwidth}
 \centering
 \vglue -40pt
 \includegraphics[width=8cm]{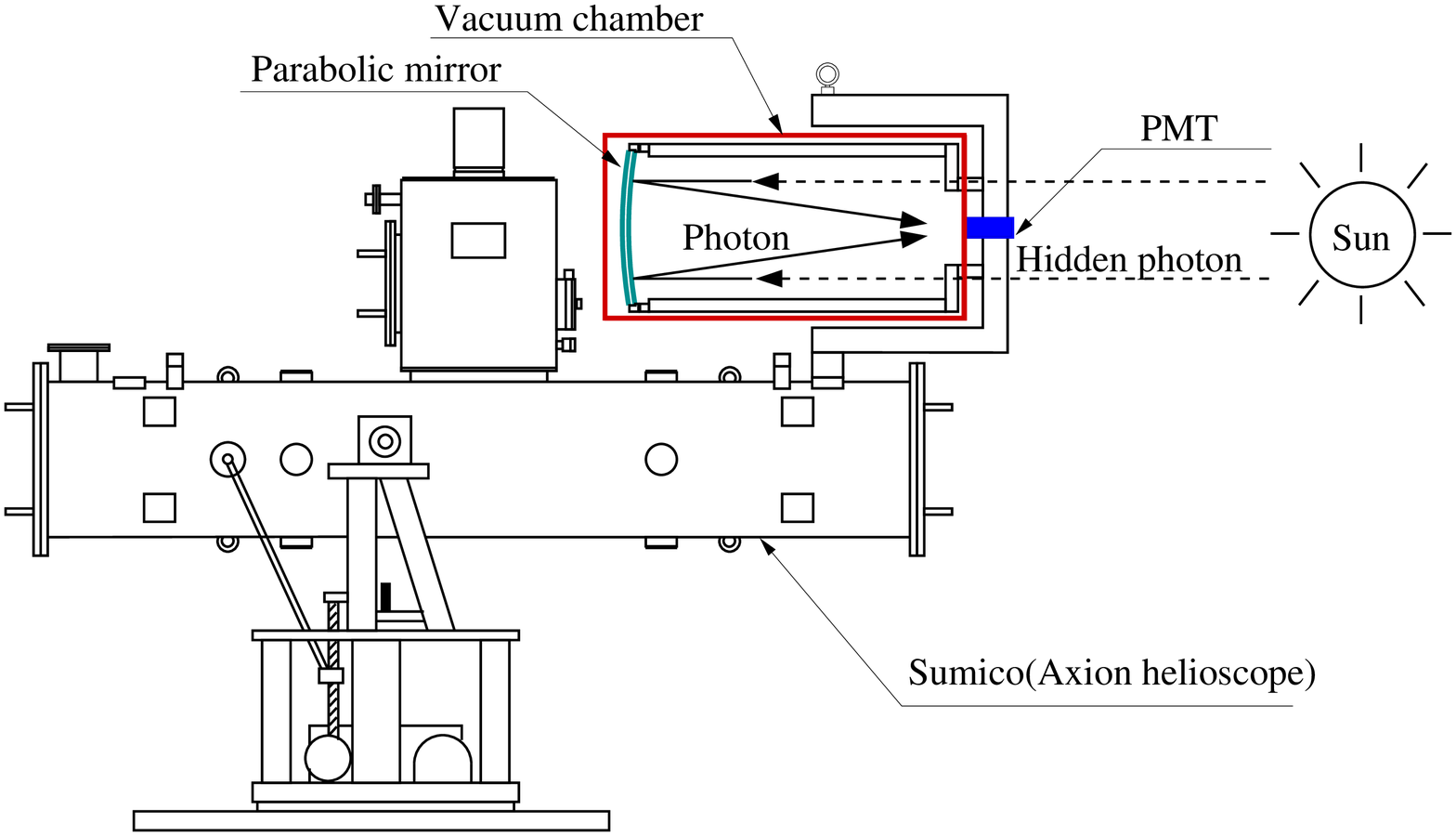}
 \caption{The schematic view of Sumico and the solar hidden photon detector.}
 \label{fig:apparatus}
\end{wrapfigure}

Our solar hidden photon detector was built as an add-on unit
mounted on Sumico, the Tokyo axion helioscope,
as shown in Fig.~\ref{fig:apparatus}.
Sumico can track the sun
with a driving range from $-28^\circ$ to $28^\circ$
in altitudinal direction and $360^\circ$ in azimuth.
The overall tracking accuracy is better than 0.5\,mrad in both directions,
which is negligibly small for this measurement.

The solar hidden photon detector consists of a vacuum chamber,
a parabolic mirror and a photomultiplier tube (PMT).
The vacuum chamber holds the conversion region in vacuum 
to keep the hidden photon to photon conversion probability high enough. 
It is a cylinder made of 1.5-mm thick stainless steel plates 
with wrinkles on its side for the mechanical reinforcement. 
The inner diameter of the vacuum chamber is 567mm and its length 
is 1200mm. 
The parabolic mirror was used to collect the conversion photons to the PMT.
It is made of aluminium deposited soda glass.
Its diameter is 500\,mm, the focal length is 1007\,mm
and the focal spot diameter is 1.5\,mm.
The reflectance measured by the manufacturer
is higher than 80\% over the wavelength range 300--650\,nm.
The mirror and its supporting structure is constructed inside the
vacuum chamber.

A low noise head-on type 25-mm diameter photon counting PMT,
Hamamatsu Photonics R3550P,
was used as the photon detector at the focal point.
It is sensitive to photons of wavelength range 
300--650\,nm with a peak quantum efficiency of 17\,\%. 
It is set at the atmospheric pressure side
and is viewing the reflected photons through a quartz glass window.
Single- and multi photon events detected by the PMT make current pulses 
which enter a charge-sensitive preamplifier (ORTEC 113) 
and a shaping amplifier (ORTEC 572).
The signal is then digitized by an ADC (Laboratory Equipment Corp. 2201A)
and pulse height spectra are
taken by a multichannel analyser (MCA).

The temperatures of the PMT and the vacuum chamber are
measured by Pt100 thermometers.
The inner pressure of the vacuum chamber is
measured by a vacuum gauge (Balzers PKR250).
During the solar tracking- and background measurements, 
the inner pressure of the vacuum chamber was lower than
$(5\pm2)\times10^{-3}\rm\,Pa$.
The effect of this residual gas on the conversion probability is negligible.

\section{Measurement and analysis}
\begin{figure}[b]
  \centerline{(a)
    \vtop{\hrule height 0pt width 0pt
      \hbox{\includegraphics[width=6cm]{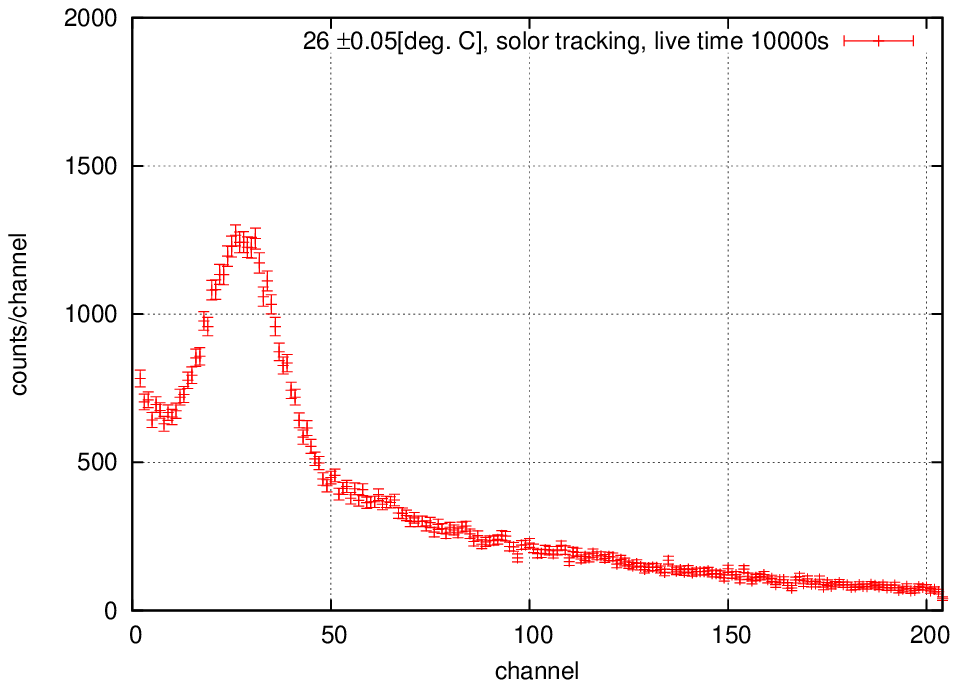}}}
    \quad
    (b)
    \vtop{\hrule height 0pt width 0pt
      \hbox{\includegraphics[width=6cm]{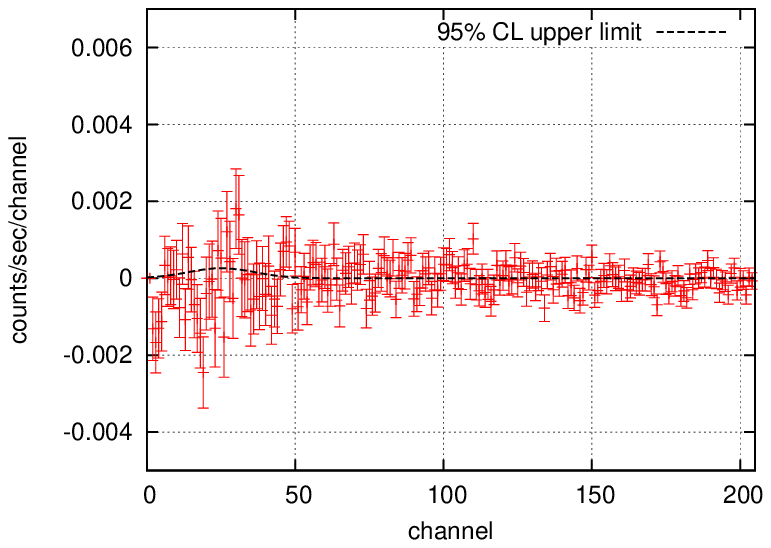}}}}
  \caption{
    (a) An example of a pulse height spectrum
    (PMT temperature $26.0\pm0.05^\circ\rm C$, solar tracking data). 
    (b) Total residual spectrum and the 95\% confidence level upper limit.
  }
  \label{fig:spec}
\end{figure}

If a hidden photon is converted into a photon in the vacuum chamber,
it would be detected by the PMT as a single photon event.
Before starting the measurement, 
the shape of a single photon spectrum in the MCA was measured
by illuminating the PMT
with a blue LED with sufficiently low current pulses.
It was fitted by a Gaussian function
which was later used as the template for the single photon analysis.

Measurements were done from October 26, 2010 till November 16, 2010. 
The solar tracking measurements were done around the time of sunrise and sunset
with tracking time of about 5 hours each.
Background was measured while the detector was directed away from the sun.

To find out the possible evidence of solar hidden photons,
the background spectrum was subtracted from the solar tracking spectrum. 
We must eliminate some systematic effects which have nothing to do with 
the solar hidden photons.
It is well known that the dark count rate gets lower as time passes
after an operating voltage is applied.
We, therefore, waited for four days until the time dependence
on the dark count rate became negligible. 

\begin{figure}[t]
  \centerline{\includegraphics[width=100mm]{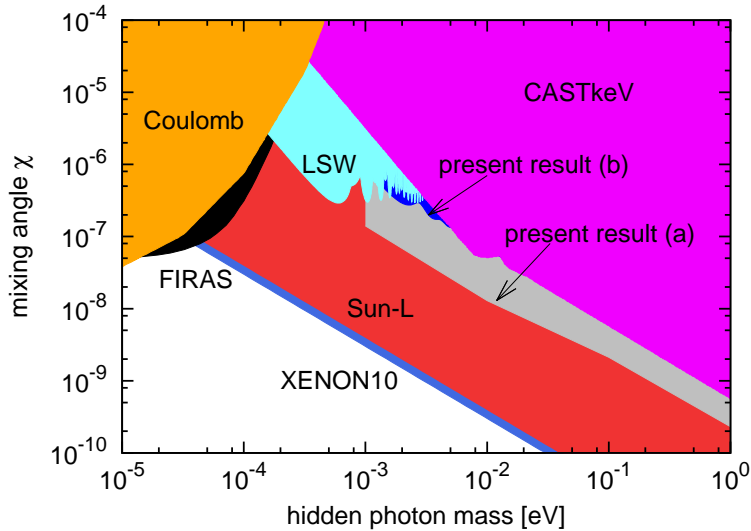}}
  \caption{
    95\,\% Confidence level upper limits to the mixing parameter $\chi$
    set by this experiment.
    Present result (a) is obtained based on the
    the newer solar hidden photon flux calculation~\cite{Redondo2, Redondo3}
    and (b) on the older conservative estimation~\cite{Redondo1}.
    Excluded area by other experiments are;
    Coulomb: tests of Coulomb's law~\cite{Clmb1,Clmb2},
    LSW: Light Shining through Walls experiments~\cite{LSW1,LSW2,LSW3,LSW4},
    CAST keV: the CAST experiment~\cite{Redondo1},
    FIRAS: FIRAS CMB spectrum~\cite{CMB},
    Sun-L: the solar luminosity constraints
    in the longitudinal channel~\cite{An1, Redondo4},
    and XENON10: interpretation of the XENON10 result in view of the
    longitudinal solar hidden photons~\cite{An2}.
  }
  \label{fig:exclusion}
\end{figure}

To avoid the systematic effect from a temperature dependence
of the dark count rate,
we subtracted background isothermally.
First, we grouped the solar tracking- and background-spectra
each with 100\,s of live time into 
21 temperature bins of 0.1$^\circ$C interval
whose central values ranging 25.9--27.9$^\circ$C.
Then, we apply the background subtraction within every temperature bin and 
obtained 21 residual spectra.
Finally, the total residual spectrum was obtained by combining them.

In the above procedure, we used only the data during the holidays
when the air conditioning system was switched off,
because we observed abrupt room temperature changes on weekdays 
due to automatic switching of the air conditioning system of the building.

Figure~\ref{fig:spec} (a) shows the pulse height spectrum
from solar tracking data
in the temperature bin $26.0\pm0.05^\circ\rm C$, as an example.
Figure~\ref{fig:spec} (b) shows the total residual spectrum.
It is worth noting that the peak from single photon events
is already evident in the background spectrum
and it vanished away after the background subtraction.

We then estimated how many single photons could there be
in the total residual spectrum
by fitting the magnitude of the Gaussian template function to it.
The best fit was obtained with
\[
  N_{\rm fit} = (-7.9\pm 6.5({\rm stat.})\pm 3.4({\rm sys.}))
  \times 10^{-3}[{\rm s}^{-1}].
\]
The systematic errors considered include an effect of Cherenkov light
emitted in the quartz glass vacuum window and the PMT window by cosmic muons,
the finite bin width of the PMT temperatures,
and the residual drift of the dark count rate after four days.

As we observed no excess of the single photon events,
the 95\% confidence level upper limit to the hidden photon counting rate
was estimated
taking the statistical and systematic errors into account:
\[
  N_{\rm UL95} = 1.02\times 10^{-2}{\rm s}^{-1} .
  \label{obs}
\]
The upper limit $N_{\rm UL95}$ is now compared with the count rate
expected by the hidden photon model with given parameters.
The 95\% confidence level upper limit to the mixing parameter $\chi$
as a function of the hidden photon mass $m_{\gamma '}$ is calculated
as shown in Fig.~\ref{fig:exclusion}.
For the solar hidden photon flux ${{\rm d}\Phi_{\rm T}\over{\rm d}\omega}$,
two cases were assumed.
One is based on the conservative estimation~\cite{Redondo1}
which is indicated by `present result (b)' in Fig.~\ref{fig:exclusion}.
The other is based on the newer estimation~\cite{Redondo2, Redondo3} 
including the contribution from a thin resonant region below the photosphere,
which is indicated by (a).
Limits set by other experiments are also shown.

\section{Conclusion and prospects}
We have searched for solar hidden photons in the visible photon
energy range using a dedicated detector for the first time. 
The detector was attached to the Tokyo axion helioscope, Sumico
which has a mechanism to track the sun.
No evidence of the existence of hidden photons
is observed in the measurement
and we set a limit on photon--hidden-photon mixing parameter $\chi$ 
depending on the hidden photon mass $m_{\gamma '}$. 
The present result improved the existing limits given by the LSW
experiments and the CAST experiment in the hidden photon mass
region between $10^{-3}$ and 1\,eV.
With recent new calculations of the longitudinal-mode hidden photon,
more stringent limits came out by the solar luminosity consideration
and also by the re-analysis of the XENON10 data.
This result is already published in Ref.~\cite{Sumico4}.

If solar tracking- and background measurements could be switched
in less than 10 minutes, the effect of the dark-count-rate drift
including that caused by the temperature changes would become negligible.
This can be achieved by slightly switching the helioscope axis
as much as 2 degrees while Sumico is tracking the sun.
A test has demonstrated that Sumico can switch between
the on-axis solar tracking
and off-axis tracking within 20 seconds
with enough direction accuracy.
Use of a parabolic mirror of larger area and
a photodetector with lower noise and higher quantum efficiency
will also improve the sensitivity.
However, such upgrades would not improve the sensitivity in terms of $\chi$
by orders of magnitude.

\section*{Acknowledgements}
This research is supported by the Grant-in-Aid for challenging Exploratory Research
by MEXT, Japan,
and by the Research Center for the Early Universe, School of Science, the University of Tokyo.


\begin{thebibliography}{99}
\bibitem{PQ1} R.~Peccei, H.~Quinn,
  Phys.\ Rev.\ Lett.\ {\bf 38}, 1440 (1977).
\bibitem{PQ2} R.~Peccei, H.~Quinn,
  Phys.\ Rev.\ D {\bf 16}, 1791 (1977).
\bibitem{Sumico} R.~Ohta, {\it et al.},
  Nucl.\ Instr.\ Meth.\ A {\bf 670}, 73 (2012).
\bibitem{Sumico1} S.~Moriyama, {\it et al.},
  Phys.\ Lett.\ B {\bf 434}, 147 (1998).
\bibitem{Sumico2} Y.~Inoue, {\it et al.},
  Phys.\ Lett.\ B {\bf 536}, 18 (2002).
\bibitem{Sumico3} Y.~Inoue, {\it et al.},
  Phys.\ Lett.\ B {\bf 668} 93 (2008).
\bibitem{Ringwald} M.~Goodsell and A.~Ringwald, Fortsch.\ Phys.\ {\bf 58}, 716 (2010).
\bibitem{Okun1} L.~Okun,
  Sov.\ Phys.\ JETP {\bf 56}, 502 (1982).
\bibitem{Holdom} B.~Holdom,
  Phys.\ Lett.\ B {\bf 166}, 196 (1986).
\bibitem{Foot} R.~Foot, X.~He,
  Phys.\ Lett.\ B {\bf 267}, 509 (1991).
\bibitem{Redondo1} J.~Redondo,
  JCAP {\bf 07}, 008 (2008) [arXiv:0801.1527 [hep-ph]],
  S.~N.~Gninenko and J.~Redondo,
  Phys.\ Lett.\ B {\bf 664}, 180 (2008)
  [arXiv:0804.3736v1 [hep-ex]].
\bibitem{Redondo2} D.~Cadamuro and J.~Redondo,
  arXiv:1010.4689v1 [hep-ph].
\bibitem{Redondo3} J.~Redondo,
  arXiv:1202.4932v1 [hep-ph].
\bibitem{An1} H.~An, M.~Pospelov, J.~Pradler,
  arXiv:1302.3884 [hep-ph].
\bibitem{Redondo4} J.~Redondo, G.~Raffelt,
  arXiv:1305.2920 [hep-ph]. 
\bibitem{An2} H.~An, M.~Pospelov, J.~Pradler,
  arXiv:1304.3461 [hep-ph].
\bibitem{JaeckelRingwald} J.~Jaeckel, A.~Ringwald,
  Annu. Rev. Nucl. Part. Sci. 2010.60:405
  [arXiv:1002.0329v1 [hep-ph]].
\bibitem{Clmb1} E.~Williams, J.~Faller, H.~A.~Hill,
  Phys.\ Rev.\ Lett.\ {\bf 26}, 721 (1971).
\bibitem{Clmb2} D.~F.~Bartlett and S.~Loegl,
  Phys.\ Rev.\ Lett.\ {\bf 61}, 2285 (1988).
\bibitem{LSW1} K.~Ehret {\it et al.} [ALPS Collaboration],
  Phys.\ Lett.\ B {\bf 689}, 149 (2010)
  [arXiv:1004 .1313v1 [hep-ex]].
\bibitem{LSW2} M.~Fouche {\it et al.} [BMV Collaboration],
  Phys.\ Rev.\ D {\bf 78}, 032013 (2008)
  [arXiv:0808.2800v1 [hep-ex]].
\bibitem{LSW3} M.~Ahlers {\it et al.}, 
  Phys.\ Rev.\ D {\bf 77}, 095001 (2008)
  [arXiv:0711.4991v1 [hep-ph]],
  A.~S.~Chou {\it et al.} [GammeV  Collaboration],
  Phys.\ Rev.\ Lett.\ {\bf 100}, 080402 (2008)
  [arXiv:0710.3783 [hep-ex]].
\bibitem{LSW4} A.~Afanasev {\it et al.} [LIPSS Collaboration],
  Phys.\ Lett.\ B {\bf 679}, 317 (2009)
  [arXiv:0810.4189v1 [hep-ex]].
\bibitem{CMB} A.~Mirizzi, J.~Redondo and G.~Sigl,
  JCAP {\bf 03}, 026 (2009)
  [arXiv:0901.0014[hep-ph]].
\bibitem{Sumico4} T.~Mizumoto {\it et al.},
  JCAP {\bf 07}, 013 (2013)
  [arXiv:1302.1000v3 [astro-ph.SR]].
\end{thebibliography}
\end{document}